\documentclass[conference,11pt]{IEEEtran}
\usepackage{amsfonts,amsmath,amssymb,url,graphicx,epsfig}

\usepackage[english]{babel}
\usepackage[ruled,vlined]{algorithm2e}
\usepackage{algpseudocode}
\usepackage{mathrsfs}
\usepackage{pifont}
\usepackage{array}
\usepackage{tabularx,ragged2e,booktabs,caption}
\usepackage{multirow}
\graphicspath{{./figs/}}
\usepackage{etoolbox}
\usepackage{blindtext}
\newcommand{\subparagraph}{}
\usepackage{titlesec}
\usepackage{textcomp}
\usepackage{xcolor}
\usepackage{flushend}
\usepackage{amssymb}
\usepackage{multirow}
\usepackage{subfig} 
\usepackage{caption}
\usepackage[
backend=biber,
style=ieee,
sorting=none
]{biblatex}
\usepackage{svg}
\graphicspath{{images/}}
\usepackage{breqn}
\usepackage{geometry}
\IEEEoverridecommandlockouts 
\begin{document}
\title{Graph Neural Network Based Access Point Selection for Cell-Free Massive MIMO Systems}
\author{ \IEEEauthorblockN{Vismika Ranasinghe, Nandana~Rajatheva,  and Matti Latva-aho}

\thanks{This work was supported by the Academy of Finland 6Genesis Flagship (grant no. 318927).}

\IEEEauthorblockA{Centre for Wireless Communications,~~University of Oulu, Finland\\
E-mail: vismika.maduka@oulu.fi, nandana.rajatheva@oulu.fi, matti.latva-aho@oulu.fi}
}
\maketitle
\thispagestyle{empty}

 
\begin{abstract}
A graph neural network (GNN) based access point (AP) selection algorithm for cell-free massive multiple-input multiple-output (MIMO) systems is proposed. Two graphs, a homogeneous graph which includes only AP nodes representing the structure of the APs in the network, and a heterogeneous graph which includes both AP nodes and user equipment (UE) nodes are constructed to represent a cell-free massive MIMO network. A GNN based on the inductive graph learning framework GraphSAGE is used to obtain the embeddings which are then used to predict the links between the nodes. The numerical results show that compared to the proximity-based AP selection algorithms, the proposed GNN based algorithm predicts the potential APs with more accuracy. Compared to the large scale fading coefficient based AP selection algorithms, the proposed algorithm does not require measured and sorted signal strengths of all the neighbouring APs. Furthermore, the proposed algorithm is scalable in terms of the number of users in the cell-free system.



\emph{Index  Terms} - AP selection, cell-free massive MIMO, graph neural network
\end{abstract}

\section{Introduction}


For 5\textsuperscript{th} generation (5G) communication  and beyond, cell-free massive multiple input multiple output (MIMO)  has attracted the attention of the research community due to its higher spectral efficiency compared to the cellular counterpart \cite{cellFreeVsSmallCells,cellFreeEmilMakingCompetitive}. This higher spectral efficiency results from a small number of user equipment being jointly served by a large number of geographically distributed access points (AP) using the same time-frequency resources. This difference is more significant when APs are densely deployed because inter-cell interference in cellular networks is leveraged as diversity gain in cell-free massive MIMO.  

For cell-free massive MIMO systems to be practical, AP selection, pilot assignment, and cluster formation algorithms need to be scalable. In \cite{scalableCellFreeEmil}, authors have presented a definition for a scalable cell-free massive MIMO system along with a scalable cell-free massive MIMO framework. Furthermore, they have proposed a scalable joint AP selection, pilot assignment and cluster formation algorithm that ensures each UE is served by at least one AP. For each UE in the network, the AP with the strongest large scale fading coefficient is selected as the master AP and then a pilot which results in minimal pilot contamination is assigned by the master AP. Depending on whether the neighbouring APs are vacant on the same pilot that is assigned to the UE by the master AP, they decide whether to serve the UE or not. The distributed nature of the algorithm ensures scalability without any burden on the fronthaul links which connect APs and central processing unit (CPU). In \cite{effectiveChannelGainAccess}, a centralized AP selection algorithm based on a metric called effective channel quality is presented. Here, the metric effective channel quality takes in to account both the channel strength and the interference. Due to the centralized nature of the algorithm, it does not fulfil all the scalability criteria presented in \cite{scalableCellFreeEmil}. In \cite{greenEnergyEfficiencyAP} and \cite{graphColoringPilot}, for each UE, APs with the strongest large scale coefficients are selected before optimizing the cell-free system for energy efficiency and pilot contamination, respectively.

One key practical limitation of choosing the APs with strongest links is the accurate measurement of large scale fading coefficients for all the APs in the network. In 3GPP specifications, among others, measurement quantity reference signal received power (RSRP) is defined to measure channel quality \cite{3gppMeasurement}. In a densely deployed cell-free network, measuring RSRP values of all the APs, sorting them, and selecting the APs with largest RSRP values, as proposed in \cite{greenEnergyEfficiencyAP} and \cite{graphColoringPilot}, can introduce unwanted delays to initial access and mobility management procedures as AP selection is included in those procedures.  As proposed in \cite{scalableCellFreeEmil}, selecting the closest APs to the master AP may require lesser number of RSRP measurements with no sorting. 
However, this AP selection may not be the optimal selection due to shadow fading. For a given deployment, the shadow fading can be fixed as it depends on the static features of the environment like buildings, hills, etc. In this work we aim to leverage this information and propose a graph neural network (GNN) based AP selection algorithm where a trained GNN is used to predict the APs with potential links once the RSRP measurements of few known set of APs are available to the neural network as the input. As the shadow fading value map is fixed in a given environment a GNN can be used to learn a set of embeddings to predict the potential links between APs and UEs.


The remainder of the paper is organized as follows. In section II, a brief introduction to GNNs is given while the system model and the GNN model we propose are presented in section III. Finally, a discussion along with the simulation results and a conclusion are given in section IV and section V, respectively.

\section{Graph Neural Networks}

Graph neural networks are used to learn a representation of a graph and perform machine learning tasks by taking the learned representation as the input. Given a set of input features per node or/and edge, graph convolution, a generalized version of the euclidean convolution, is performed to obtain a set of embeddings. These can then be used for machine learning tasks like link prediction, node clustering, or node classification.  There are two types of convolutions that can be performed on graphs. One is based on spectral graph theory and is called spectral convolutions while the other one, which is based on the edges connecting the nodes, is called spatial convolutions. Since our model uses only spatial convolutions, in this section we limit our discussion to GNNs with only spatial convolutions.

 Formally, the graph $\mathcal{G}$ is defined by $\mathcal{G} = (\mathcal{V}, \mathcal{E})$ where $\mathcal{V}$ is the set of nodes and $\mathcal{E}$ is the set of edges between the nodes. An edge going from node $u \in \mathcal{V}$ to node $v \in \mathcal{V}$ is denoted by $(u,v) \in \mathcal{E}$. In the case of a heterogeneous graph, we can partition the nodes in to disjoint sets $\mathcal{V}_1, \mathcal{V}_2 \dots \mathcal{V}_k$ where $\mathcal{V} = \mathcal{V}_1 \cup \mathcal{V}_2 \cup \dots \cup \mathcal{V}_k$ and $\mathcal{V}_i \cap \mathcal{V}_j = \varnothing, \forall i \neq j$. Since the edges connecting the different types of nodes represent different relations, edge notation is also extended to include the edge type as $(u,v,\tau) \in \mathcal{E}$ where $\tau$ is the edge type connecting $u \in \mathcal{V}_i$ to $v \in \mathcal{V}_j$. Furthermore, we can represent each type of edge with a matrix called \textit{adjacency matrix} where $\mathrm{A}_\tau[u,v] = 1$ if $(u,v,\tau) \in \mathcal{E}$ and $\mathrm{A}_\tau[u,v] = 0$ otherwise. When the edges of the graph are all undirected then the adjacency matrix becomes symmetric. In some graphs, edges can have weights where adjacency matrix will have arbitrary real values rather than $\lbrace0,1\rbrace$.
 
 
 In order to incorporate domain specific information into a graph, nodes are represented with feature vectors. In this work we denote a set of feature vectors using $\mathrm{X} \in \mathbb{R}^{d\times\vert \mathcal{V}\vert}$, where we assume the ordering of the features of the nodes is consistent with the ordering of the nodes in the adjacency matrix. Here each node is represented with a $d$ dimensional vector $x_i$ where $i \in \mathcal{V}$. Given the input graph $\mathcal{G}$ and the node feature set $\mathrm{X}$, a GNN based on message passing framework \cite{graphbook} can be used to obtain the set of node embeddings $z_u, \forall u \in \mathcal{V}$  for downstream machine learning tasks. During each message passing iteration, the \textit{hidden embedding} ${h}_u^{(n)}$ corresponding to each node $u \in \mathcal{V}$ and iteration $n$ is updated using the spatial convolution given by,
 \begin{equation}
      \begin{split}
      {h}_u^{(n)} = \text{UPDATE}^{(n)}\Big({h}_u^{(n-1)}, \text{AGGREGATE}^{(n)} \big(\\{h}_v^{(n-1)}, \forall v \in \mathcal{N}(u)\big) \Big),
      \label{spatialConvo}
 \end{split}
 \end{equation}

where UPDATE and AGGREGATE are arbitrary differentiable functions and  $\mathcal{N}(u) = \lbrace v \in \mathcal{V} : (v,u) \in \mathcal{E} \rbrace$, i.e. the set of neighbours of the node $u$. The set of hidden embeddings after $n$ iterations are denoted by $\mathrm{H}^{(n)} \in \mathbb{R}^{d\times\vert \mathcal{V}\vert}$. In the first iteration, $h^{(0)}_u = x_u$ and after $N$ iterations we can use the final embeddings $z_u = h^{(N)}_u$ as a input to the downstream machine learning tasks.



\section{Graph neural network based access point selection}

\begin{figure*}[htbp]
  \centering
  \includegraphics[width=1\textwidth]{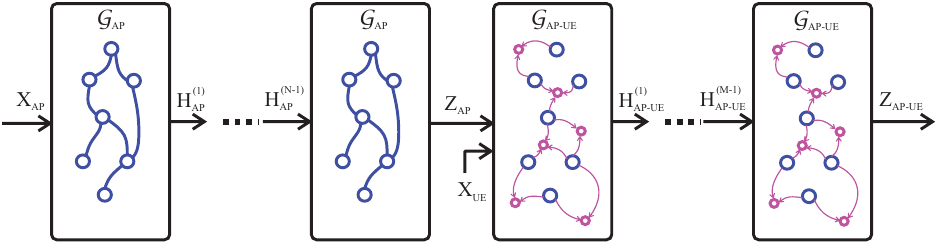}
  \caption{GNN architecture based on GraphSAGE \cite{Graphsage} framework}
  \label{graphSAGE}
\end{figure*}

\subsection{System Model}

We consider a cell-free system consisting of $L$ APs and $K_t$ UEs which are trying to establish a connection with the cell-free network at the time instance $t$. In order to predict the links between the AP nodes and the UE nodes we consider two graphs where first one is formed using only the $L$ AP nodes and  second one is formed using both $L$ AP nodes and $K_t$ UE nodes. Since the algorithm can be repeated for each time instance, henceforth, we denote the number of UEs in the graph as $K$. Furthermore, the set of APs and the set of UEs are represented by $\mathcal{L}$ and $\mathcal{K}$, respectively.

\subsection{Graph Construction}
\label{graphconstruction}
\subsubsection{AP graph}

In a given scenario, since the AP nodes are fixed in the environment, it is important that the GNN is able to learn the structure of the AP placement. For a given AP node $l \in \mathcal{L}$, distance to the other neighbouring AP nodes are calculated and edges are created between the closest $c_{AP}$ AP nodes and the AP node $l$. The set of closest $c_{AP}$ APs to AP $l$ is denoted by $\mathcal{C}_l$. The feature vector $x_l^{AP} \in \mathbb{R}^L$ for AP node $l$ is  $x_l^{AP}(\hat{l}) = R_{l,\hat{l}} \text{ if } l \in \mathcal{C}_{ \hat{l}} \text{ otherwise } x_l^{AP}(\hat{l}) = 0$, i.e. if the AP $l$ is in the set of closest $c_{AP}$ APs to the neighbouring AP $\hat{l}$, only then $x_l^{AP}(\hat{l}) = R_{l,\hat{l}}$.  Here, $ R_{l,\hat{l}}$ is the strength of the signal transmitted from AP $l$ and measured at AP $\hat{l}$. Here, the sparsity of the AP feature vector depends on the density neighbouring APs. The above step is repeated for $\forall l \in \mathcal{L}$ forming the AP graph and the feature set $\mathrm{X}_{AP}$ for AP nodes. Finally all the edges between APs are made undirected such that the adjacency matrix $\mathrm{A}_{AP}$ becomes symmetric. This graph is denoted by $\mathcal{G}_{AP} = (\mathcal{L}, \mathcal{E}_{AP})$ where $\mathcal{E}_{AP}$ is the set of bidirectional edges between AP nodes.

\subsubsection{AP-UE graph}

Similar to the AP selection algorithm proposed in \cite{scalableCellFreeEmil}, UE $k \in \mathcal{K}$ first establish a connection with the AP $l_k \in \mathcal{L}$ which has the strongest link to the UE $k$. This is known as the \textit{master AP} of the UE $k$. For the link prediction task, only the closest $c_{UE}$ APs to the AP $l_k$  are considered and directional edges from those APs to the UE $k$ are created. This initial clustering based on physical distances or measured signal strengths is done to avoid the class imbalance between positive edges and negative edges in classification task. Henceforth,  this initial set of APs is denoted by $\mathcal{C}^{cluster}_{l_k}$. After the initial connection with the master AP $l_k$, it requests the UE $k$ to measure and report RSRPs for $\hat{c}_{UE}$ closest APs to the master AP, where $\hat{c}_{UE} \ll c_{UE}$. This set of APs for which the RSRP measurements are requested by master AP is denoted by $\mathcal{C}^{R}_{l_k}$. The feature vector $x_k^{UE} \in \mathbb{R}^L$ for the UE node $k$ is $x_k^{UE}(\hat{l}) = R_{\hat{l},k} \text{ if } \hat{l} \in \mathcal{C}^{R}_{l_k} \text{ otherwise } x_l^{UE}(\hat{l}) = 0$. Finally, this step is repeated for all $\forall k \in \mathcal{K}$ forming the AP-UE graph and the feature set $\mathrm{X}_{UE}$ for UE nodes. This graph is denoted by $\mathcal{G}_{AP-UE} = (\mathcal{L} \cup \mathcal{K}, \mathcal{E}_{UE})$ where $\mathcal{E}_{UE}$ is the set of directional edges from AP nodes to UE nodes.

\subsection{Graph Neural Network Architecture}

In order to learn the representation of $\mathcal{G}_{AP}$ and $\mathcal{G}_{AP-UE}$, we use GraphSAGE framework proposed in \cite{Graphsage}. GraphSAGE is a graph representation learning framework which can be used for inductive applications where the graphs are dynamic. 


As illustrated in Figure \ref{graphSAGE}, $N$ spatial graph convolutions, based on  (\ref{spatialConvo}), are performed on  $\mathcal{G}_{AP}$ with the feature set $\mathrm{X}_{AP}$ as the input. Then $M$ spatial convolutions are performed on $\mathcal{G}_{AP-UE}$, where $\mathrm{Z}_{AP}$ will be the input feature  set of the AP nodes and  $\mathrm{X}_{UE}$ will be the input feature set of the UE nodes. The directional edges between AP nodes and UE nodes ensure that the resulting AP embeddings are independent from the UE nodes in the graph, as messages are only passed from AP nodes to UE nodes. In order to match the distributions of the inputs $\mathrm{Z}_{AP}$ and $\mathrm{X}_{UE}$ to the $\mathcal{G}_{AP-UE}$, $\mathrm{X}_{UE}$ is passed through a single layer feed-forward network. From the multiple aggregator functions proposed in \cite{Graphsage}, we use the pooling aggregator, for node $k \in \mathcal{L} \cup \mathcal{K}$,  given by 

\vspace{-0.5cm}
\begin{multline}
    \text{AGGREGATE}^{(n)} = \max\Big(\Psi_n(h_v^{(n)}), \\  \forall v \in \mathcal{N}(k)\Big),
\end{multline}
 where $\Psi_n$ represent the two layer feed-forward network for the $n^{\text{th}}$ iteration. UPDATE function for the node $k \in \mathcal{L} \cup \mathcal{K}$ is given by
 
 \begin{multline}
    \text{UPDATE}^{(n)} = \mathrm{SELU} \Big( \mathbf{W^{(n)}}h_k^{(n)} + \\ \text{AGGREGATE}^{(n)} \big(h_v^{(n)}, \forall v \in \mathcal{N}(k)\big) \Big),
\end{multline}

where $\mathbf{W^{(n)}}$ is the trainable parameter matrix for the $n^{\text{th}}$ iteration and SELU stands for scaled exponential linear unit.

\subsection{Training and Inference}
\label{trainingandinference}

In the training phase, after obtaining the final node embeddings $\mathrm{Z}_{AP-UE}$, the confidence of the link between UE $k \in \mathcal{K}$ and  AP $l \in \mathcal{C}^{cluster}_{l_k}$ is calculated using

\begin{equation}
     S_{l,k} = \sigma(\mathrm{z}_l^T\mathrm{z}_k),
\end{equation}

 where $\sigma$ is the sigmoid function and $\mathcal{C}^{cluster}_{l_k}$ is the initial cluster of APs, closest to master AP $l_k$, considered for the link prediction task. In order to train the GNN, training graphs with known positive and negative edges are provided. For the UE $k \in \mathcal{K}$, the links $(l,k) \in \mathcal{E}_{UE}$, $\forall l \in \mathcal{C}^{cluster}_{l_k}$ where $R_{l_k,k} - R_{l,k} < D_{dB}$, are considered as potential links and hence labeled as positive edges in the graph  $\mathcal{G}_{AP-UE}$.
  The links $(l,k) \in \mathcal{E}_{UE}$, $\forall l \in \mathcal{C}^{cluster}_{l_k}$ which does not meet the above criterion is labeled as negative edges. Here $R_{l,k}$ is the measured RSRP of the AP $l$ reported by the UE $k$ and $D_{dB}$ is a hyper parameter that needs be decided when creating the training dataset. The assumption behind this criterion is that there exist a closed loop power control between the master AP $l_k$ and the UE $k$. Therefore, selecting a fixed number of APs with strongest links is not effective as signal strength of some of the links may be significantly weaker than the strongest link.  After the edges are labeled, the GNN is trained using binary cross entropy as the loss function.
 
 In the inference stage, for the UE $\hat{k}$ that established a connection with the AP $l_{\hat{k}}$, as explained above, $\mathcal{G}_{AP-UE}$ is constructed after obtaining $R_{l,\hat{k}}, \text{ }\forall l \in \mathcal{C}^{R}_{l_{\hat{k}}}$ from measurement reports which are configured through RRC signaling. Then the embeddings for the nodes are obtained to calculate the confidence values and classify the links. Since the AP node embeddings are independent from $X_{UE}$ and the number of UE nodes, final node embeddings for the AP nodes can be cached to save computational power.

\section{Experimental Results and Analysis}

\subsection{Parameters and Setup}

For numerical simulation of the proposed algorithm, we considered cell-free networks with $L=50 \text{ and } 100$ APs, which are independently and uniformly distributed in a $2 \text{km} \times 2 \text{km}$  and  $1 \text{km} \times 1\text{km}$ squares, respectively. In order to simulate the measured RSRP value between AP $l$ and UE $k$, we adopted the path loss model used in \cite{scalableCellFreeEmil}, which is given by

\begin{equation}
    R_{l,k} = \Upsilon - 10\alpha\log_{10}\Big(\frac{d_{l,k}}{1 \text{km}}\Big) + F_{l,k},
    \label{RSRPsim}
\end{equation}

where $d_{l,k} \text{ }[km]$ is the distance between the AP $l$ and the UE $k$, $\alpha = 3.76$ is the path loss exponent, $\Upsilon = -35.3 \text{dB}$ is the median channel gain at a reference distance of 1 km, and $F_{l,k}$ is the shadow fading. Here $F_{l,k}$ is obtained using correlated fading maps generated for each AP based on the algorithm proposed in \cite{FadingCorrelated}. 

To construct a training graph, as explained in section \ref{graphconstruction}, $K_{train} = 100$ UEs are independently and uniformly placed in the map, then the RSRP values are calculated according to (\ref{RSRPsim}), and subsequently the links are labeled as positive or negative edges as described in section \ref{trainingandinference}. It was empirically observed that by setting $D_{dB} = 10$, uplink power received by all the APs with positively labeled links represent on average 85\% of the total uplink power received by all the APs in the network.  100 such training graphs are used to train the network with batch size 1.  For validation, the number of UEs ($K_{validate}$) in each graph is made random to simulate a dynamic demand for the AP selection algorithm. As described in section \ref{graphconstruction}, the graph construction parameters along with GNN hyper parameters are tabulated in Table \ref{pramsCons}. In the GNN,  hidden embeddings after every message passing iteration are represented using $L$ dimensional vectors. For the implementation of the proposed GNN based algorithm,  PyTorch Geometric \cite{geometricpaper}, the geometric learning extension of the open source machine learning framework  PyTorch \cite{pytorch} was used.

\begin{table}[]
\begin{tabular}{|p{1.1cm}|p{4.9cm}|p{1cm}|}
\hline
Parameter & Description & Value \\ \hline
   $c_{AP}$    & Number of closest APs considered to create edges between APs             &   5    \\ \hline
    $c_{UE}$      & Number of APs closest to the master AP considered for link prediction             &   10, 20    \\ \hline
     $\hat{c}_{UE}$     & Number of APs closest to the master AP for which the RSRP measurements are performed             &  2     \\ \hline
     $\mathrm{N}$     & Number of message passing iterations for $\mathcal{G}_{AP}$             &  2     \\ \hline
     $\mathrm{M}$     & Number of message passing iterations for $\mathcal{G}_{AP-UE}$             &  2     \\ \hline
\end{tabular}

\caption{Graph construction parameters}
\label{pramsCons}
\end{table}

\begin{figure}[t]
  \includegraphics[width=8.5cm, height=15cm]{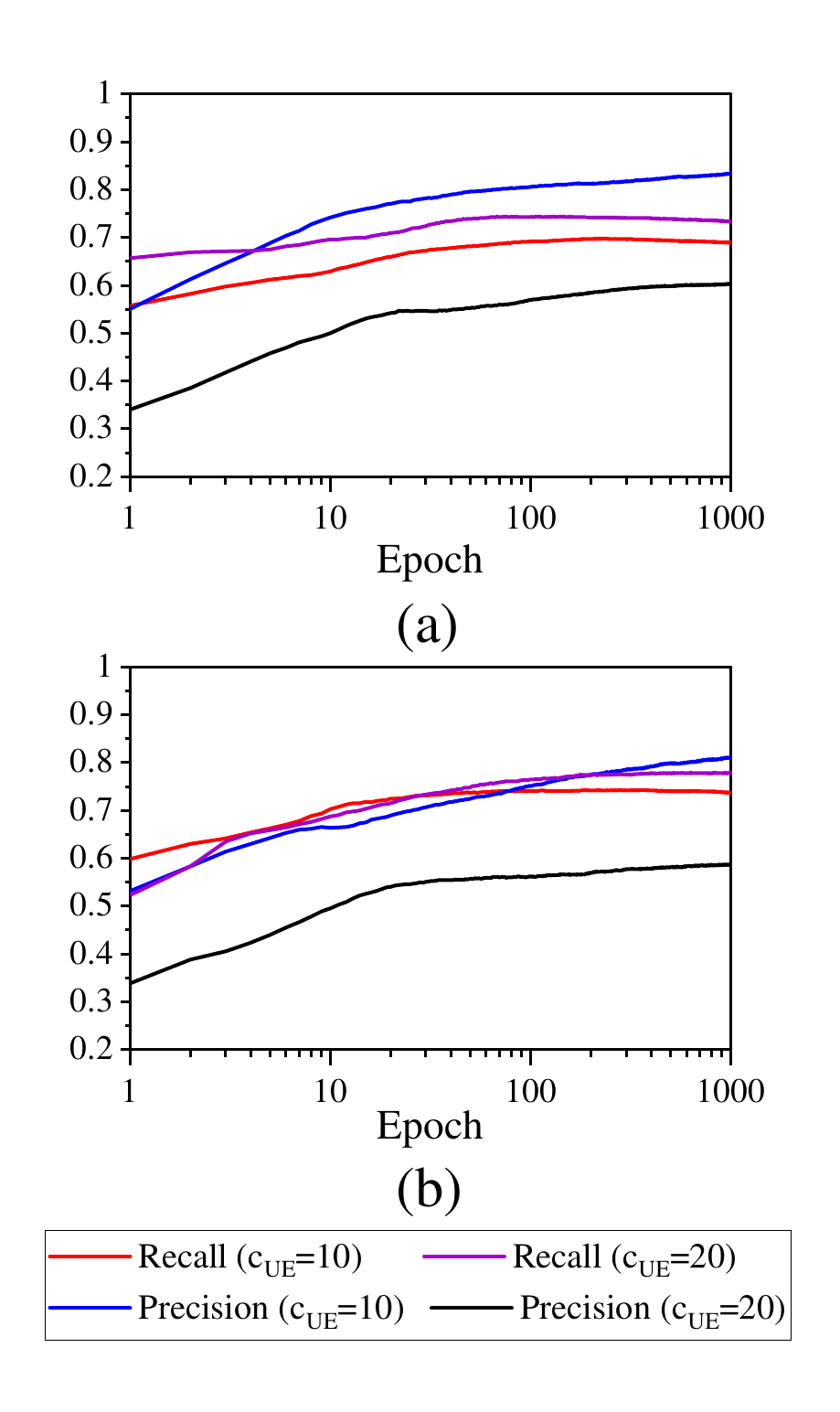}
  \caption{Precision and Recall when $L=100$ [(a)] and $L=50$ [(b)] }
  \label{results_precisionRecall}

\end{figure}

\begin{figure}[t]
  \includegraphics[width=8.5cm, height=15cm]{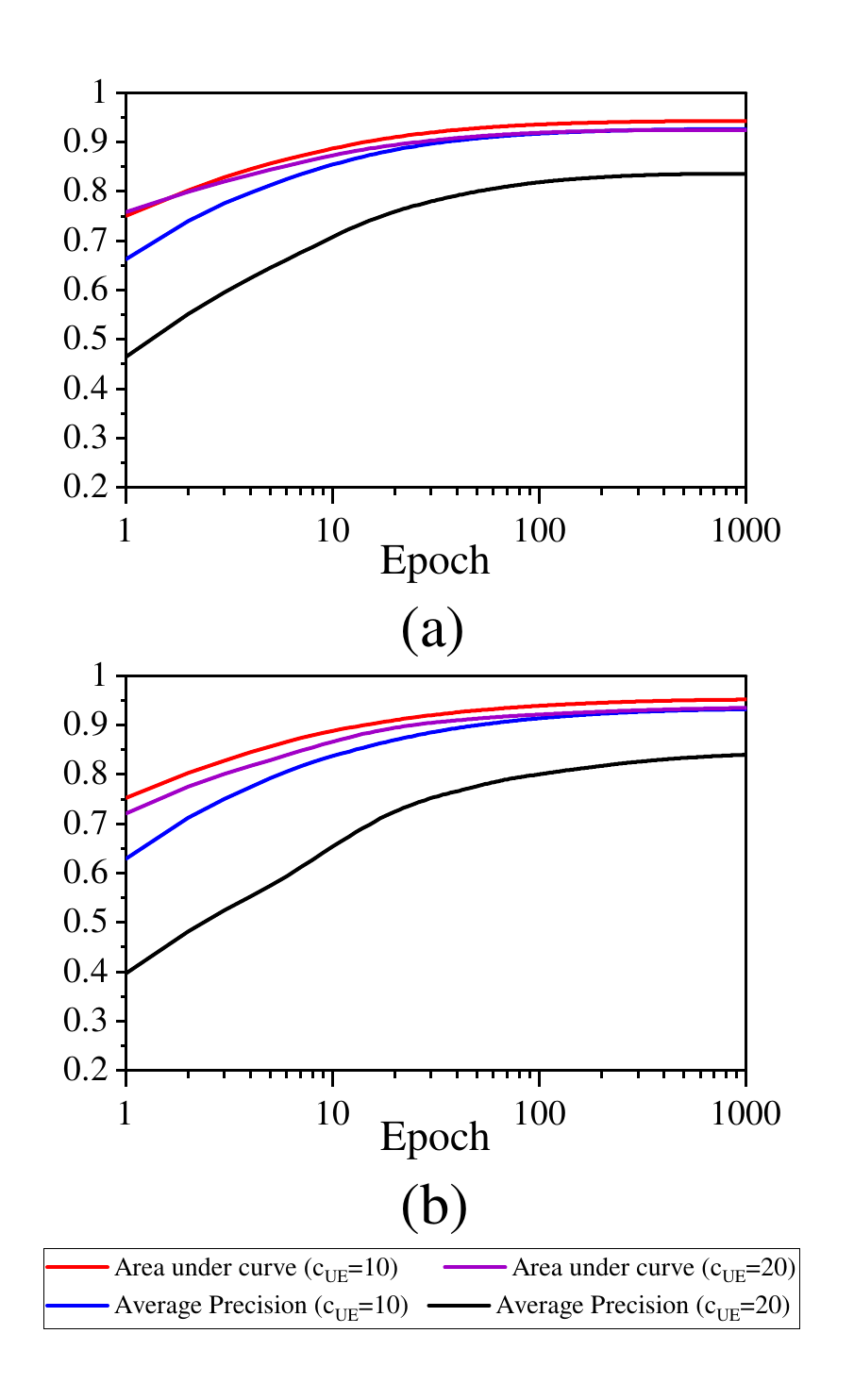}
  \caption{Area under curve and average precision when $L=100$ [(a)] and $L=50$ [(b)] }
  \label{results_aucap}

\end{figure}

Figure \ref{results_precisionRecall} illustrates precision and recall for the two cell-free networks with $L=100$ and $L=50$. Here, we used the threshold 0.5 to classify the edges on the graph in order to calculate the precision and recall values. When only the 10 APs closest to master AP ($c_{UE}=10$) are considered for the AP selection, in each cell-free network with $L=100$ and $L=50$ about 16\% and 13\% positive links are ignored, respectively. When $c_{UE}$ is increased to 20, respectively, these values drop to 9\% and 6\%. Hence, the maximum recall that can be achieved by the GNN is limited by the resulting recall after this initial clustering procedure. From Figure \ref{results_precisionRecall}(a), it can be observed that when  $L=100$ and $c_{UE} = 10$ proposed method achieves a precision and a recall up to 0.83 and 0.68 respectively. This recall reflects the recall due to GNN's inaccuracy and the recall due to initial clustering. When $c_{UE} = 20$, it can be observed that recall is improved slightly at the expense of precision. This is due to fact that as we increase $c_{UE}$, the number of APs considered for link prediction, we increase the negative edges in the graph resulting a class imbalance between positive and negative edges. From Figure \ref{results_precisionRecall}(b), similar effects  on precision and recall can be observed when $c_{UE}$ is increased in the simulation of the cell-free network with $L=50$.

In Figure \ref{results_aucap}, we present two additional binary classification evaluation metrics area under the curve (AUC) and average precision. Both AUC and average precision metrics are able to show the discriminative  capabilities of the proposed model independent of the decision threshold. From Figure \ref{results_aucap}, it can be observed when $c_{UE}=10$ AUC is above 0.9 and average precision is above 0.8 for both simulations with $L=100$ and $L=50$. When increasing $c_{UE}$, it can be observed that average precision is affected significantly. With regard to the scalability of the proposed GNN method, as the feature vectors of the APs and the UEs are represented using $L$ dimensional vectors, increasing the number of APs increases the complexity of the GNN model. As a result, the proposed algorithm is scalable only when the number of UEs in the network is increased. 

From the generated validation data sets for $L=100$, it was observed that on average only 33\% of the closest 3 APs to the master AP have a positive link to the UE that meets the criterion described in section \ref{trainingandinference}. Furthermore, the probability that the closest AP to the master AP having a positive link to the UE is on average 42\%. Compared to the metrics of the proximity based algorithms presented in \cite{scalableCellFreeEmil} our algorithm achieves a precision and a recall up to 0.83 and 0.68. Hence, it can be concluded that compared to proximity based AP selection algorithms, our proposed method is capable of predicting the APs with potential links accurately with limited number of RSRP measurements as the only input. Furthermore, from the presented results, it can be observed that the difference in AP densities of the two networks (25 APs per 1 $\text{km}^2$ and 50 APs per 1 $\text{km}^2$) doesn't affect the performance of the proposed algorithm. 

\section{Conclusions}

In this work, a GNN based AP selection algorithm is proposed. Compared to the AP selection based on signal strength, which requires a large number of RSRP measurements and sorting them, proposed algorithm only requires a small number of RSRP measurements. The proposed method is capable of predicting the potential links to the UE, up to a precision and a recall of 0.83 and 0.68, respectively, based on three RSRP measurements including the RSRP of the AP with the strongest link to the UE. Furthermore, results show that the proposed GNN architecture outperforms proximity based AP selection algorithms as shadowing significantly affects the accuracy of proximity based AP selection algorithms. Compared to selecting APs based on the APs proximity to the master AP, proposed algorithm leverage the knowledge of the static correlated fading value map and AP placement to predict the APs with potential links. Even though the proposed algorithm is scalable as the number of UEs in the network increases, scalability in terms of number of APs is limited compared to the proximity based distributed AP selection algorithm proposed in \cite{scalableCellFreeEmil}.

\IEEEpubidadjcol 

\renewcommand*{\bibfont}{\footnotesize}

\printbibliography

\end{document}